\documentclass{article}
\usepackage{spconf,graphicx}
\usepackage{amsmath,amssymb,amsfonts,bm}
\usepackage{hyperref}
\usepackage{cite}
\usepackage{color}
\usepackage{array}
\usepackage{algorithmic}
\usepackage{graphicx}
\usepackage{textcomp}
\usepackage{xcolor}
\usepackage{booktabs}
\usepackage{hyperref}
\usepackage{makecell}
\usepackage{multirow}
\usepackage[caption=false]{subfig}
\usepackage{float}
\usepackage{hyperref}
\usepackage{bbding}
\usepackage{pifont}
\title{First-Shot Unsupervised Anomalous Sound Detection With Unknown Anomalies Estimated by Metadata-Assisted Audio Generation}

\name{{Hejing Zhang$^{1,2}$,  Qiaoxi Zhu$^3$, Jian Guan$^{1,2*}$, Haohe Liu$^{4}$, Feiyang Xiao$^{1,2}$, Jiantong Tian$^{1,2}$,}
\thanks{*Corresponding author.}
\thanks{This work was partly supported by the GHfund under Grant No. 202302026860.}
}{Xinhao Mei$^4$, Xubo Liu$^{4}$, Wenwu Wang$^4$}
\address{
     $^1$ Group of Intelligent Signal Processing, College of Computer Science and Technology,\\  Harbin Engineering University, Harbin, China\\
     $^2$ National Engineering Laboratory for Modeling and Emulation in E-Government, \\ Harbin Engineering University, Harbin, China\\
     $^3$ Centre for Audio, Acoustics and Vibration, University of Technology Sydney, Ultimo, Australia\\
     $^4$ Centre for Vision Speech and Signal Processing, University of Surrey, Guildford, UK\\}
\begin{document}
%\ninept
%
\maketitle
\begin{abstract}
First-shot (FS) unsupervised anomalous sound detection (ASD) is a brand-new task introduced in DCASE 2023 Challenge Task 2, where the anomalous sounds for the target machine types are unseen in training. Existing methods often rely on the availability of normal and abnormal sound data from the target machines.  
However, due to the lack of anomalous sound data for the target machine types, it becomes challenging when adapting the existing ASD methods to the first-shot task. In this paper, we propose a new framework for the first-shot unsupervised ASD,  where metadata-assisted audio generation is used to estimate unknown anomalies, by utilising the available machine information (i.e., metadata and sound data) to fine-tune a text-to-audio generation model for generating the anomalous sounds that contain unique acoustic characteristics accounting for each different machine type. We then use the method of Time-Weighted Frequency domain audio Representation with Gaussian Mixture Model (TWFR-GMM) as the backbone to achieve the first-shot unsupervised ASD. Our proposed FS-TWFR-GMM method achieves competitive performance amongst top systems in  DCASE 2023 Challenge Task 2, while requiring only 1\% model parameters for detection, as validated in our experiments.

\end{abstract}

\section{Introduction}
\label{sec:intro}

Anomalous sound detection (ASD) aims to distinguish between the normal and anomalous operating states of a machine based on the sounds emitted from it \cite{ Koizumi2020, Dohi2023, zeng2022robust, liu2022anomalous, guan2023time}. However, due to the infrequent occurrence and potential diversity of anomalous sound, it is challenging and time-consuming to gather sufficient training data for anomalous sound covering various situations. To mitigate this issue, unsupervised ASD, utilising only normal sounds during training, becomes a desirable, albeit challenging, option.

First-shot (FS) unsupervised ASD has been introduced for the Detection and Classification of Acoustic Scenes and Events (DCASE) 2023 Challenge Task 2 \cite{Dohi2023, harada2023first}, aiming to detect target machine types' anomalous sounds that are unseen in training. There are three sets of information used for training: (1) anomalous and normal sounds from the \textit{reference} machine types, (2) normal sounds from the \textit{target} machine types, (3) metadata, including machine type and attributes on operational and environmental conditions, as the label of each of the above sounds. These target machine types (e.g., Vacuum, ToyTank, ToyNscale, ToyDrone, Bandsaw, Grinder, and Shaker) are entirely distinct from the reference machine types (e.g., Fan, Gearbox, Bearing, Slider, ToyCar, ToyTrain, and Valve). State-of-the-art ASD methods often rely on the availability of normal and abnormal data from the target machines. However, in practice, anomalous sound data for the target machine types may be difficult to capture due to their rare occurrence in practice. This makes it difficult to adapt these existing ASD methods to the first-shot task, as discussed by the DCASE 2023 Challenge Task 2 organisers \cite{Dohi2022-2, Dohi2023}.

To address this challenge, we present a new framework for the first-shot unsupervised ASD with unknown anomalies estimated by metadata-assisted audio generation. Specifically, we use a text-to-audio (TTA) generation model for synthesizing anomalous and normal sounds for the target machine type. We use the state-of-the-art TTA model, i.e. AudioLDM  \cite{liu2023audioldm-key}, but fine-tuned using all the available data in the first-shot scenario, including the anomalous and normal sounds from the reference machine types, normal sounds from the target machine types, and their corresponding metadata describing the operational and environmental conditions of these machines. The proposed approach is built on the ASD model in our previous study \cite{guan2023time}, which is a Time-Weighted Frequency domain audio Representation (TWFR) with Gaussian Mixture Model (GMM). For this reason, we abbreviate the proposed first-shot approach as FS-TWFR-GMM. In this model, we use a hyperparameter $r$ to highlight the important information of the audio representation in the time domain that may differ between machine types. It is determined for the target machine type by fine-tuning using synthesised normal and anomalous sounds rather than using real normal and anomalous sounds as in \cite{guan2023time}. 

\begin{figure}[t]
    \centering
    \includegraphics[width=0.8\columnwidth]{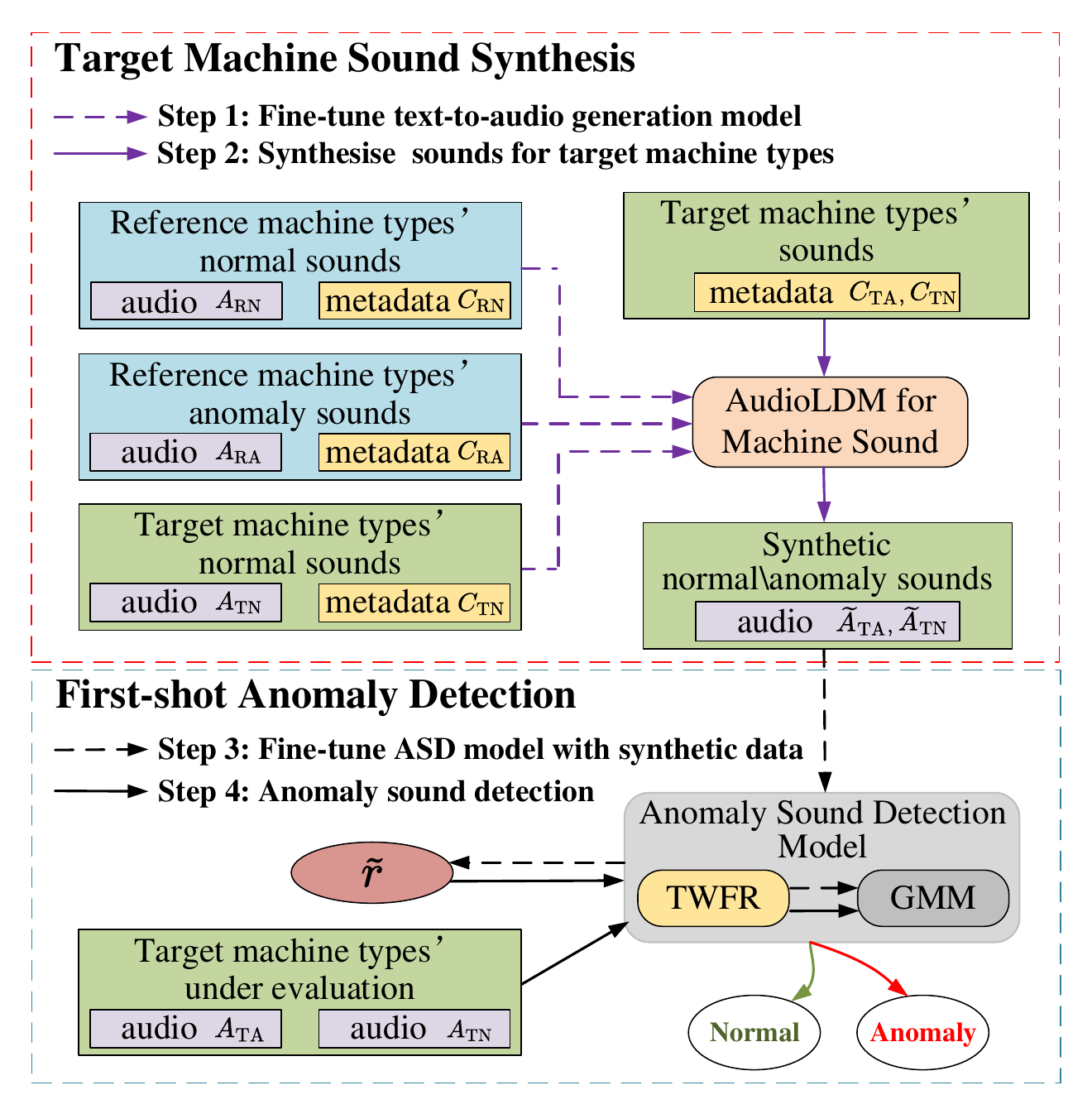}
    \vspace{-4mm}
     \caption{The proposed first-shot unsupervised ASD method using TWFR-GMM as the backbone. For target machine sound synthesis (steps 1-2), we use normal and anomalous sounds from reference machine type(s) and normal sounds from the target machine type(s), along with corresponding metadata detailing machine operating and monitoring conditions. For first-shot anomaly detection (step 3-4), we identify the unseen anomalous sounds from the target machine type(s).}
    \vspace{-0.4cm}
    \label{fig:workflow}
\end{figure}

Our method is the first to estimate unknown anomalies in unsupervised ASD using machine information for audio generation. Existing anomaly synthesis methods (e.g., \cite{WilkinghoffFKIE2023, MAE_Data_Aug,inoue2020detection, chen2023effective}) only use normal sound data, but not abnormal sounds of the reference machines or metadata. In contrast, our method exploits all available audio data and metadata, giving improved quality in the synthesised machine sounds, which captures the features of actual sounds from the corresponding machine type. Furthermore, our approach is versatile in accommodating a wide range of machine metadata, regardless of type and format, thus alleviating challenges posed by diverse machine attributes and label formats frequently encountered in real-world scenarios. 

Experiments show that the proposed method provides competitive performance for the first-shot unsupervised ASD. The preliminary version \cite{tianfirst} forms the core of the system ranked the 7th in DCASE 2023 Challenge Task 2. Our improved version in this paper is now between 3rd and 4th place, with only $1.34\%$ in the AUC metric and $2.27\%$ in the pAUC metric lower than the top method. Notably, our approach requires vastly reduced resources due to a non-deep-learning design, i.e., only $1.1\%$ of the number of parameters as used in the top method in DCASE 2023 Challenge. Thus, the proposed method is promising for practical applications with computing resource constraints.

\section{Proposed Method}
\label{sec:method}

Fig.~\ref{fig:workflow} outlines the proposed FS-TWFR-GMM method. First, we synthesize the sounds of target machine types with a fine-tuned text-to-audio generation model, utilising normal and anomalous sounds and their metadata from the reference machine types and normal sound data from the target machine type, as detailed in Section~\ref{ssec:method_generation}. Then, the ASD model tuned for each target machine type with the synthetic machine sounds is used to detect the unseen anomalous sounds, as detailed in Section~\ref{ssec:method_r}. Note that the backbone ASD model TWFR-GMM can be replaced with other ASD models, such as \cite{chen2023effective, zeng2022robust, mai2022explaining}. 

\subsection{Metadata-Assisted Machine Sound Synthesis}
\label{ssec:method_generation}

\subsubsection{Metadata based Machine Sound Captioning}

In the training dataset for ASD, the label of machine sounds contains related metadata, e.g., machine operating status. In our method, we generate captions based on the metadata, for example, using captions to describe the machine's operating status. We then use captions as textual prompt, and generate synthetic sound using a TTA model. First, we convert the metadata to captions as 
\begin{equation}
\label{eq:Convert}
     c = \bm{F}_{c}(l)
\end{equation}
where $\bm{F}_{c}(\cdot)$ denotes the captioning function. It converts the label $l$ of a machine sound to caption ${c}$, with a predefined descriptive text template for each different machine type, with examples illustrated in Table ~\ref{tab:metadata}.

\begin{table*}[!htbp]
\label{tab:metadata}
\scriptsize
    \centering
        \caption{Examples of available metadata within the descriptive text captions for machine sounds from audio labels, including the normal and anomalous sound of the reference machine type, i.e., ToyCar, and the normal sound of the target machine type, i.e., Grinder. }
        \resizebox{\textwidth}{!}{
            \begin{tabular}{ccccccccc}
			\toprule
			&\Large{Machine type} &\Large{Example of the label (metadata)} &\Large{Caption for text-to-audio generation}\\
                \midrule
                &\Large{\textbf{ToyCar}} &\Large{section\_00\_source\_test\_\textbf{\textit{normal}}\_0001\_car\_\textbf{\textit{B2}}\_spd\_\textbf{\textit{31V}}\_mic\_\textbf{\textit{1}}.wav} &\Large{This is the \textbf{\textit{normal}} sound of a \textbf{\textit{toy car}} with model \textbf{\textit{B2}} and speed \textbf{\textit{31V}}, recorded by a microphone placed at the position \textbf{\textit{1}}.}\\
                &\Large{\textbf{ToyCar}} &\Large{section\_00\_source\_test\_\textbf{\textit{anomaly}}\_0001\_car\_\textbf{\textit{B2}}\_spd\_\textbf{\textit{31V}}\_mic\_\textbf{\textit{1}}.wav} &\Large{This is the \textbf{\textit{anomaly}} sound of a \textbf{\textit{toy car}} with model \textbf{\textit{B2}} and speed \textbf{\textit{31V}}, recorded by a microphone placed at the position \textbf{\textit{1}}.}\\
                &\Large{\textbf{Grinder}} &\Large{section\_00\_source\_train\_\textbf{\textit{normal}}\_0000\_grindstone\_\textbf{\textit{2}}\_plate\_\textbf{\textit{2}}.wav} &\Large{This is the \textbf{\textit{normal}} sound of a \textbf{\textit{grinding}} machine with grindstones \textbf{\textit{2}} and metal plates \textbf{\textit{2}}.}\\
                \bottomrule
            \end{tabular}
        }
   \vspace{-0.28cm}
\label{tab:metadata}
\end{table*}

\subsubsection{Fine-tuning AudioLDM for Machine Sound Synthesis}
We use the AudioLDM algorithm to synthesise machine sounds related to a specific target machine type in terms of the captions generated in Section 2.1.1. The AudioLDM algorithm uses the contrastive language-audio pretraining (CLAP) \cite{elizalde2023clap} to build a shared latent space between text embeddings of the captions and audio embeddings of the sounds, and uses the latent diffusion model (LDM) \cite{liu2023audioldm-key} on a continuous audio representation for text-to-audio generation, conditioned on the caption. To tailor this model for our task, we fine-tune a pre-trained AudioLDM \cite{liu2023audioldm-key}, using the machine sound and caption pairs, as follows,
\begin{equation}
\label{eq:fine-tune}
   \mathcal G {\leftarrow}  P (  \bm{A}|{ \bm{C}} )
\end{equation}
where $P$ denotes the pre-trained AudioLDM model, and $\mathcal G$ is the AudioLDM model fine-tuned by a set of machine audios $\bm{A}$, with the corresponding set of captions $\bm{C}$ as the condition.

\begin{equation}
\label{eq:a-and-c}
     \left\{\begin{matrix}\bm{A} = \{ {{\bm A}}_\text{RN}, {{\bm A}}_\text{RA}, {{\bm A}}_\text{TN} \}
     \\{\bm C} = \{ {\bm C}_\text{RN}, {\bm C}_\text{RA}, {\bm C}_\text{TN} \} 
    \end{matrix}\right.
 \end{equation}
where $\bm{A}_\text{RN}$, $\bm{A}_\text{RA}$ and $\bm{A}_\text{TN}$  respectively represent the sets of audio signals for the reference machine type's normal sounds, the reference machine type's anomalous sounds, and the target machine type's normal sounds, with corresponding sets of captions $\bm{C}_\text{RN}$, $\bm{C}_\text{RA}$, and $\bm{C}_\text{TN}$.

To synthesise sounds for the target machine types, we use the corresponding captions as the condition and gradually denoise from the fine-tuned LDM distribution to estimate the true data distribution and generate audio, that
\begin{equation}
\label{eq:generation}
    \left\{\begin{matrix}\widetilde{{\bm A}} _\text{TN} = \mathcal G ({\bm C} _\text{TN} )
    \\\widetilde{{\bm A}} _\text{TA}  = \mathcal G ({\bm C} _\text{TA})
    \end{matrix}\right.
\end{equation}
where $\widetilde{{\bm A}} _\text{TN}$ and $\widetilde{{\bm A}} _\text{TA}$ denote the sets of  synthetic audios for target machine types' normal sounds and anomalous sounds, respectively. 

In first-shot unsupervised ASD, the audios and captions for target machine types' anomalous sounds do not exist in the training stage. Therefore, we obtain the captions set ${\bm C} _\text{TA}$ for target machine types' anomalous sound generation by replacing the word ``normal'' in ${\bm C} _\text{TN}$ with ``anomaly''.

\subsection{First-Shot Unsupervised ASD Using Synthesised Sounds }
\label{ssec:method_r}

With the synthetic normal and anomaly sounds accounting for characteristics of specific target machine types, we can train the ASD model for the first-shot scenario by optimising audio feature representations to distinguish between normal and abnormal sounds effectively. In this paper, we adapt TWFR-GMM for the first-shot scenario, resulting in FS-TWFR-GMM. 

The TWFR-GMM algorithm \cite{guan2023time} {obtains the time-weighted frequency domain audio representation (TWFR) $R(\mathbf{X}) \in  \mathbb{R}^{M}$, by incorporating a hyperparameter $r$ for each machine type, as follows}
\begin{equation}
\label{eq:gwrp}
    R(\mathbf{X}) = Ranking(\mathbf{X})\cdot \left [ \frac{r^0}{z(r)}, \frac{r^1}{z(r)}, ..., \frac{r^{N-1}}{z(r)}  \right ]^\top
\end{equation}
where $\mathbf{X} \in \mathbb{R}^{M  \times N}$ is the log-mel spectrogram of an audio signal with $M$ mel-bins and $N$ time frames. $Ranking (\cdot)$ denotes the operation of re-arranging $\mathbf{X}$ in descending order for the energy values over time frames for time weight calculation, following \cite{guan2023time}. Here, $r$ determines the weight assigned to each time frame, and $z(r) = {\textstyle \sum_{n=1}^{N}} r^{n-1}$ is for weight normalisation. $\top$ denotes the transposition operation.

The audio representation $R(\cdot)$ is trained using normal sounds, but fine-tuned with anomalous sounds for each machine type's dynamic and unique sound characteristics, which is then used for audio feature extraction in the detection stage with GMM to achieve anomaly detection in \cite{guan2023time}. However, it is not applicable in the first-shot scenarios, as there are no abnormal sounds existing for the target machine types.

In this paper, to adapt TWFR-GMM for the first-shot scenario, the hyperparameter $r$ is estimated by optimising the following cost with the synthesised machine sounds,
\begin{equation}
\label{eq:r-obteain}
\widetilde{r} = \underset{r}{\text{argmax}} \left \{ E( r, \widetilde{{\bm A}}_{\text{TA}} , \widetilde{{\bm A}}_{\text{TN}}) \right \}
\end{equation}
where $\widetilde{r}$ denotes the estimated value of $r$, and $E(\cdot)$ is evaluation metric for ASD following \cite{guan2023time}. We set the selection range $r \in \left [ 0, 1.10 \right ] $, and the selection interval is $0.01$. When $r$ exceeds one, it gives greater weights to the time frames with lower energy, whereas when $r$ is less than one, it gives higher weights to time frames with higher energy. This approach considers the diverse audio patterns observed among different types of machines. For instance, some machines produce loud anomalous sounds, while others exhibit short-term stalls caused by extraneous object interference, as discussed in \cite{koizumi2019toyadmos}.

The synthesised normal and anomalous sounds ($\widetilde{{\bm A}}_{\text{TN}}$ and $\widetilde{{\bm A}}_{\text{TA}}$) are employed to optimise the hyperparameter $r$ in TWFR. In contrast, the GMM uses real normal sounds of the target machine type (${\Large a}_{\text{TN}}$). This approach helps minimise the potential bias introduced by the sounds generated by AudioLDM. Other detailed implementations of TWFR can be found in \cite{guan2023time}. 

\begin{table*}[t]
    \centering
    \caption{Performance comparison with DCASE 2023 Challenge Task 2 top submissions.}
    \resizebox{\textwidth}{!}{
    \begin{tabular}{cccccccccccccccccc}
        \toprule
        \toprule
        \multirow{2}{*}{Method} & \multirow{2}{*}{ Ranking} & \multicolumn{2}{c}{ToyDrone} & \multicolumn{2}{c}{ToyNscale} & \multicolumn{2}{c}{ToyTank} & \multicolumn{2}{c}{Vacuum} & \multicolumn{2}{c}{Bandsaw} & \multicolumn{2}{c}{Grinder} & \multicolumn{2}{c}{Shaker} & \multicolumn{2}{c}{Average} \\
        \cmidrule(lr){3-4} \cmidrule(lr){5-6} \cmidrule(lr){7-8} \cmidrule(lr){9-10} \cmidrule(lr){11-12} \cmidrule(lr){13-14} \cmidrule(lr){15-16} \cmidrule(lr){17-18}
         &  & AUC & pAUC & AUC & pAUC & AUC & pAUC & AUC & pAUC & AUC & pAUC & AUC & pAUC & AUC & pAUC & AUC & pAUC \\
        \midrule
	Jie\_IESEFPT \cite{junjieanomaly}
			& 1  & 58.03  & 51.58  & {89.03}  & {77.74}  & 60.33  & {61.53}  & {96.18}  & 85.32  
                & {65.66}  & {53.35}  & {66.63}  & 62.45  & {68.08}  & {55.97}  & {69.75} & {62.03}\\
	Lv\_HUAKONG \cite{lvunsupervised}
			& 2  & 54.84  & 49.37  & 82.71  & 57.00  & 74.80  & 63.79  & 93.66  & 87.42  
                & 58.48  & 50.30  & 66.69  & 61.22  & 74.24  & 65.24  & {70.05}  & {60.11}\\
	Jiang\_THUEE \cite{jiangthuee}
			& 3  & 55.83  & 49.74  & 73.44  & 61.63  & 63.03  & 59.74  & 81.98  & 76.42
                & 71.10  & 56.64  & 62.18  & 62.41  & 75.99  & 64.68  & {68.03}  & {60.71}\\
	\textbf{FS-TWFR-GMM (Proposed)} 
			& $-$  & 56.28  & 50.89  & 64.33  & 54.16  & 62.60  & 57.47  & 82.75  & 75.84  
                & 78.31  & 61.62  & 61.75  & 54.98  & 83.39  & 71.32  & {68.41}  & {59.76}\\ 
        Wilkinghoff\_FKIE \cite{WilkinghoffFKIE2023}
               & 4  & 53.90  & 50.21  & 87.14  & 76.58  & 63.43  & 62.21  & 83.26  & 74.00  
               & 66.06  & 52.87  & 67.10  & 62.11  & 65.91  & 50.24  & {67.95} & {59.58}\\
        Guan\_HEU \cite{tianfirst}
               & 7  & 62.93  & 52.05  & 68.94  & 54.21  & 66.41  & 60.63  & 79.47  & 72.47
               & 57.22  & 50.76  & 62.38  & 54.96  & 78.46  & 61.47  & {67.12} & {57.32}\\
               \midrule
        DCASE2023\_Baseline \cite{harada2023first}
               & 9  & 58.93  & 51.42  & 50.73  & 50.89  & 57.89  & 53.84  & 86.84  & 65.32
               & 69.10  & 57.54  & 60.19  & 59.55  & 72.28  & 62.33  & 63.41  & 56.82 \\
        \bottomrule
        \bottomrule
    \end{tabular}
    }
    \vspace{-0.4cm}
    \label{tab:experiments}
\end{table*}

\begin{table}[t]
\label{tab:parameters}
\scriptsize
    \centering
        \caption{
        Comparison of the number of parameters.
        }
		\resizebox{\columnwidth}{!}{
            \begin{tabular}{cccccccccc}
			\toprule
                \toprule
			\multicolumn{2}{c}{\multirow{1}{*}{Method}} &\multicolumn{1}{c}{Ranking}&\multicolumn{1}{c}{Training (one-off cost)}&\multicolumn{1}{c}{Detection}\\
                \midrule
			\multicolumn{2}{c}{Jie\_IESEFPT \cite{junjieanomaly}} 
			& 1 & 3M  & 3M\\
			\multicolumn{2}{c}{Lv\_HUAKONG \cite{lvunsupervised}} 
	        & 2 & 300M  & 300M\\
			\multicolumn{2}{c}{Jiang\_THUEE \cite{jiangthuee}} 
			& 3 & 6M  & 6M\\
			\multicolumn{2}{c}{\textbf{FS-TWFR-GMM (Proposed)}} 
			& $-$ & 33K+792M (AudioLDM)  & \textbf{33K}\\ 
               \multicolumn{2}{c}{Wilkinghoff\_FKIE \cite{WilkinghoffFKIE2023}} 
                & 4 & 34M  & 34M\\
               % \midrule
               \multicolumn{2}{c}{Guan\_HEU \cite{tianfirst}}
                & 7 & 33K+792M (AudioLDM)  & \textbf{33K}\\
               \midrule
               \multicolumn{2}{c}{DCASE2023\_Baseline \cite{harada2023first}}
               & 9 & 267K  & 267K\\			
                \bottomrule
                \bottomrule
            \end{tabular}
        }
    \vspace{-0.4cm}
\label{tab:parameters}
\end{table}

\begin{table}[t]
\label{tab:ablation}
\scriptsize
    \centering
        \caption{Comparison of single systems, namely FS-TWFR-GMM and its initial version (System 2 of the ensemble system \cite{tianfirst}). With or without extended $r$ range refers to selecting $r$ from $[0, 1.1]$ or $[0, 1]$. With or without RS refers to removing or keeping the silence part in the synthetic machine sounds.}
		\resizebox{\columnwidth}{!}{
            \begin{tabular}{ccccccccccccc}
			\toprule
                \toprule 
			\multicolumn{2}{c}{\multirow{1}{*}{Method}} &\multicolumn{1}{c}{Extended $r$ range}&\multicolumn{1}{c}{RS}&\multicolumn{1}{c}{AUC }&\multicolumn{1}{c}{pAUC } \\
                \midrule
			\multicolumn{2}{c}{System 2 of Guan\_HEU \cite{tianfirst}}
                & \ding{55}  & \ding{55}  & 65.07  & 57.69\\
                \midrule
			\multicolumn{2}{c}{\multirow{3}{*}{\textbf{FS-TWFR-GMM}}}
                & \checkmark & \ding{55}  & 65.22  & 57.72\\
			\multicolumn{2}{c}{}
                & \ding{55} & \checkmark  & 67.90  & 59.19\\
			\multicolumn{2}{c}{}
                & \checkmark & \checkmark  & \textbf{68.41}  & \textbf{59.76} \\
                \bottomrule
                \bottomrule
            \end{tabular}
        }
    \vspace{-0.3cm} 
\label{tab:ablation}
\end{table}

\begin{table}[!htbp]
\label{tab:r-ablation}
\scriptsize
    \centering
        \caption{Comparison of using hyperparameter $r$ from no training data, generated, or real machine sound data.}
        \resizebox{\columnwidth}{!}{
            \begin{tabular}{ccccccccc}
			\toprule
                \toprule
			&\multicolumn{2}{c}{Methods} &\multicolumn{3}{c}{Training data}  &\multicolumn{1}{c}{AUC} &\multicolumn{1}{c}{pAUC} &\multicolumn{1}{c}{Average} \\
                \midrule
			&\multicolumn{2}{c}{$r=0$} &\multicolumn{3}{c}{None}  & 56.44 & 54.04
                & 55.24 \\
			&\multicolumn{2}{c}{$r=1$} &\multicolumn{3}{c}{None}  & 66.78 &\textbf{60.28}
                & 63.53 \\
			&\multicolumn{2}{c}{\textbf{FS-TWFR-GMM}} &\multicolumn{3}{c}{Synthetic}
                 & \textbf{68.41} & 59.76 & \textbf{64.08}\\
                \midrule
			&\multicolumn{2}{c}{TWFR-GMM} &\multicolumn{3}{c}{Real}
                & 71.55 & 61.62 & 66.59\\
                \bottomrule
                \bottomrule
            \end{tabular}
        }
    \vspace{-0.3cm}
\label{tab:r-ablation}
\end{table}

\begin{table}[!htbp]
\label{tab:r}
\scriptsize
    \centering
        \caption{Selected $r$ from synthetic or real machine sounds.}
        \resizebox{\columnwidth}{!}{
            \begin{tabular}{ccccccccc}
			\toprule
			\multicolumn{2}{c}{\multirow{1}{*}{\small{Training data}}} &\multicolumn{1}{c}{\small{ToyDrone}}&\multicolumn{1}{c}{\small{ToyNscale}}&\multicolumn{1}{c}{\small{ToyTank}}&\multicolumn{1}{c}{\small{Vacuum}}&\multicolumn{1}{c}{\small{Bandsaw}}&\multicolumn{1}{c}{\small{Grinder}}&\multicolumn{1}{c}{\small{Shaker}} \\
                \midrule
			\multicolumn{2}{c}{\small{Synthetic}}
                & \small{1.02} & \small{1.00} & \small{0.99} & \small{0.84} & \small{1.03} & \small{0.96} & \small{1.01} \\
            \multicolumn{2}{c}{\small{Real}}
                & \small{1.01} & \small{1.00} & \small{0.87} & \small{0.94} & \small{1.02} & \small{0.99} & \small{1.02} \\
			\multicolumn{2}{c}{\textit{\small{Difference}}}
                & \small{0.01} & \small{0.00} & \small{0.12} & \small{0.10} & \small{0.01} & \small{0.03} & \small{0.01} \\
                \bottomrule
            \end{tabular}
            }
        \vspace{-0.3cm}
\label{tab:r}
\end{table}

\section{Experiments}
\label{sec:experiments}

\subsection{Experimental Setup}

\textbf{Dataset:} We use the DCASE 2023 Challenge Task 2 dataset \cite{Dohi2023}, including seven reference machine types (Fan, Gearbox, Bearing, Slider, ToyCar, ToyTrain, Valve) and seven target machine types (Vacuum, ToyTank, ToyNscale, ToyDrone, Bandsaw, Grinder, Shaker). In the training set, for each reference machine type, there are 1100 normal sound clips and 100 abnormal sound clips, while for each target machine type, there are 1000 normal sound clips. In the evaluation set, there are 200 sound clips with unknown conditions (normal or abnormal) for each target machine type.

\noindent\textbf{Evaluation metrics:} The area under the receiver operating characteristic curve (AUC) and the partial-AUC (pAUC) are commonly used for performance evaluation \cite{Koizumi2020, dohi2021flow, liu2022anomalous, guan2023anomalous}, where pAUC represents the AUC over a low false-positive-rate range [$0, 0.1$] \cite{Koizumi2020}. A larger value indicates better anomalous sound detection performance.

\subsection{Results}
\label{ssec: results}

Tables~\ref{tab:experiments}~and~\ref{tab:parameters} show that the proposed FS-TWFR-GMM has a significant advantage in the number of parameters (33k) required for the detection stage and achieves competitive performance ranking between the 3rd and 4th places amongst top systems in the DCASE 2023 Challenge Task 2 on first-shot unsupervised ASD, with only $1.34\%$ in AUC and $2.27\%$ in pAUC lower than the 1st placed method. 

The initial version of FS-TWFR-GMM, i.e., System 2 of the ensemble system \cite{tianfirst} achieved the 7th place in DCASE 2023 Challenge Task 2. In comparison, the FS-TWFR-GMM version proposed in this paper optimises $r$ over an extended range $[0, 1.1]$, and is shown to be more effective, as shown in Table ~\ref{tab:ablation}.

Table~\ref{tab:r-ablation} shows that the proposed method fine-tuned from synthetic data is generally better than {blindly} setting the hyperparameter ($r=0$ for max pooling, or $r=1$ for average pooling), which represents the straightforward approach due to the unavailability of anomaly data in first-shot ASD. Furthermore, the performance of the proposed unsupervised method is only $3.14\%$ in AUC and $1.94\%$ in pAUC lower than the performance achieved by the fully supervised approach employing real abnormal and normal data directly from the evaluation set to optimise $r$ in TWFR-GMM.

Table~\ref{tab:r} shows $r$ selected in terms of the synthetic sounds or real anomalous sounds in the evaluation set with nearly no difference. Moreover, it is adapted to the unique sound patterns of each machine type, through prioritising lower or higher energy time frames or treating all time frames equally ($r>1$, $r<1$, or $r=1$).

\section{Conclusion}
\label{sec:page}

We have presented a new framework for the first-shot unsupervised anomalous sound detection using a text-to-audio generation model to synthesise normal and abnormal machine sounds while leveraging all available training data. With our approach, unseen anomalies in new machine types can be estimated. As a result, it becomes easier to distinguish between normal and unknown anomaly sounds. The first-shot unsupervised method FS-TWFR-GMM implements the proposed framework on the time-weighted frequency domain audio representation with the Gaussian mixture model. It performs similarly to the state-of-the-art first-shot unsupervised ASD methods. Furthermore, the proposed framework can be used with other ASD systems for the first-shot scenarios.

\vfill\pagebreak

% \bibliographystyle{IEEEbib}
\bibliographystyle{IEEEtran}
\bibliography{refs}

\end{document}